\documentclass{aipproc}
\usepackage{srcltx}
\layoutstyle{6x9}

\def\fm3{\rm ~fm^{-3}}

\begin{document}

\begin{abstract}

Models for the internal composition of Dense Compact Stars are reviewed as well
as macroscopic properties derived by observations of relativistic processes.
Modeling of pure neutron matter Neutron Stars is presented and crust properties
are studied by means of a two fluid model. 

%The internal composition of Dense Compact Stars is addressed. Compact stars are modelled according to an Equation of state that may contain Hadrons 
%(Baryons or Hyperons) or deconfined quarks in a Quark Matter phase. 
%Relativistic processes like cooling of Neutron Stars, Eddington Flux and Gravitational Wave emission allow for determination of their macroscopic properties providing insights into their interior.
%In particular the macroscopical crust properties are studied. The crust-core transition in NS is studied by means of a two fluid model leading to the determination of the bottom of the crust.

\end{abstract}

\title{Relativistic Processes and the Internal Structure of Neutron Stars}{}
\author{D. E. Alvarez-Castillo, S. Kubis \\ \it Institute of Nuclear Physics, Radzikowskiego 152, 31-342 Krak\'ow, Poland}{}

\maketitle

\section{Introduction}

Neutron stars (NS) are stellar objects that represent one possible end of
stellar evolution. They are in hydrostatic equilibrium in which the
gravitational force is balanced by the internal one resulting from the Pauli
exclusion principle. Theorized first by Landau (1932) and Zwicky (1934), Neutron
Stars have been first observed as pulsars in 1967 by Jocellyn Bell. The latest
observations of NS point out that the most extreme physical conditions are
present in them.  They are the densest compact objects ($\rho \sim
10^{14}-10^{16}$ $\textrm{g}/\textrm{cm}^{3}$), fastest rotating stars (as fast as 716 MHz as
measured)  and fastest moving objects in the galaxy ($v\sim 1083$ $\textrm{km}/\textrm{s}$). They
possess the highest magnetic fields ($B=10^{15}$ G and largest surface gravity
($10^{14}$ $\textrm{cm}/\textrm{s}^{2}$). Their interior contains superconducting material with
the highest expected temperature value ($T_{c}=10^{9}$ K),  and even neutrinos
can be trapped in proto-neutron stars with temperatures at birth of about
700,000 million K \cite{2004Sci...304..536L}.
\vspace{-.5cm}

\section{Neutron Stars Models}

Given such extreme conditions various theoretical models have been proposed to
describe the composition of dense nuclear matter  which are translated into an
Equation of State (EoS). Based on our present knowledge of nuclear interactions,
EoSs can be categorized in a general form as

%Based on our present knowledge of nuclear interactions as well as in terms of the constitutent quarks,
%the building blocks of matter described by Quantum Chromodynamics, Equations of State can be cathegorized in a general form as

\begin{itemize}
 \item (Hadronic) Pure neutron matter: mostly n, but also p, e and $\mu$ particles.
 \item (Hadronic) Hyperonic matter: n, p, hyperons($\Lambda$, $\Sigma$, $\Xi$) and leptons (e, $\mu$).
\end{itemize}
\begin{itemize}
 \item (SQM) Strange Quark Matter. Deconfined quarks, where the name goes to the heaviest quark that appears as density increases (in this case the strange quark).
%It can be that such state of matter is in color super conducting phase (CSC), in which quarks form a superconductor. Other exotic phases are also possible.
\item Boson Condensates ($\pi$, K): Bosons can be present as a Bose-Einstein condensate state of matter (BEC).
\end{itemize}

In terms of these EoSs dense compact stars can be classified as:

\begin{itemize}
 \item[-] Neutron Stars: both crust and core are described by a hadronic EoS.
 \item[-] Hybrid Stars: their crust is hadronic but having a quark matter core. 
 \item[-] Quark Stars: Only described by a Quark Matter EoS.
\end{itemize}
The macroscopic, relativistic structure of a Neutron Star is described as
follows. For a static, non-rotating star, the Einstein equations give
 the Tolman-Oppenheimer-Volkoff equations \cite{1983bhwd.book.....S}:
%\begin{equation}
% R^{\alpha \beta} -\frac{1}{2}Rg^{\alpha \beta}=\frac{8\pi G}{c^{4}}T^{\alpha\beta}
%\end{equation}
%\begin{equation}
% T^{\alpha\beta} = \left(\rho+ \frac{p}{c^{2}}\right)u^{\alpha}u^{\beta}+p g^{\alpha \beta}
%\end{equation}
%\begin{eqnarray}
%R^{\alpha \beta} -\frac{1}{2}Rg^{\alpha \beta}=\frac{8\pi G}{c^{4}}T^{\alpha\beta} & , &
%T^{\alpha\beta} = \left(\rho+ \frac{p}{c^{2}}\right)u^{\alpha}u^{\beta}+p g^{\alpha \beta}.
%\end{eqnarray}
\begin{equation}
 \frac{dp}{dr}=-\frac{(\rho+p/c^{2})G(m+4\pi r^{3}p/c^{2})}{r^{2}(1-2Gm/rc^{2})}
,~~~~~~~~~~~  \frac{dm}{dr}=4\pi r^2 \rho.
\end{equation}
%\begin{equation}
%\frac{dp}{dr}=-\frac{(\rho+p/c^{2})G(m+4\pi r^{3}p/c^{2})\Lambda}{r^{2}}
%\end{equation}
%\begin{equation}
%\frac{dm}{dr}=4\pi r^2 \rho
%\end{equation}
%\begin{equation}
%\frac{d\nu}{dr}=\frac{2G(m+4\pi r^{3}p/c^{2})\frac{1}{1-2Gm/rc^{2}}}{r^{2}}
%\end{equation}
%\begin{equation}
%\Lambda=\frac{1}{1-2Gm/rc^{2}}
%\end{equation}
This system requires an EoS of the form $p(\rho)$ and is to be solved for mass $m$, pressure $p$ and density $\rho$ inside the star. 
To find a solution it's necessary to choose a central density ($\rho_{c}$) and take into account the boundary conditions that $m(r=0)=0$, 
$m(R)=M$ and $p(r=R)=0$ where $R$ and $M$ are the total radius and total mass of the star. As for the crust, the moment of inertia $I$ and mass $M$ 
are derived, in the frame of General Relativity \cite{1994ApJ...424..846R}, as:
\begin{equation}
 I \simeq \frac{J}{1+2GJ/R^{3}c^{2}} ,~~~~~~~~~~J=\frac{8\pi}{3}\int_{0}^{R}r^{4}\left(\rho+\frac{p}{c^2}\right),
\end{equation}
\begin{equation}
\Delta I_{crust}=\frac{2}{3}(M_{crust}R^{2})\frac{1-2GI/R^{3}c^{2}}{1-2GM/Rc^{2}}
\end{equation}
%\begin{equation}
%I \simeq \frac{J}{1+2GJ/R^{3}c^{2}} 
%\end{equation}
%\begin{equation}
 %J=\frac{8\pi}{3}\int_{0}^{R}r^{4}\left(\rho+\frac{p}{c^2}\right)\Lambda dr
%\end{equation}
where $M_{crust}=M-M_{core}$ the difference between the total mass and the mass of the core. To determine the latter the crust-core transition point must 
be known, i.e. the edge of the solid crust where the liquid core starts.
\begin{figure}[t!]
\resizebox{0.8\textwidth}{8cm}
{\includegraphics{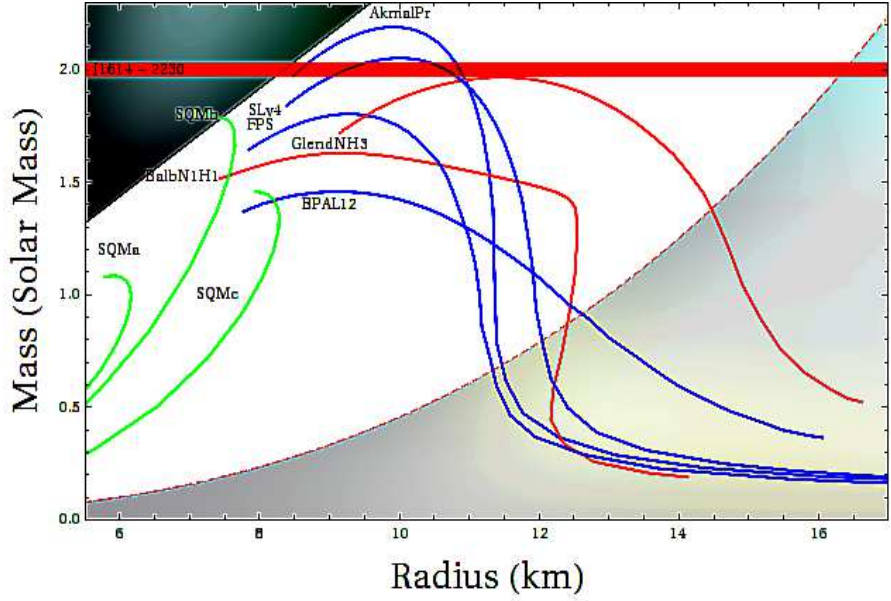}}
\caption{Mass vs Radius Relation for EoSs described in Table~\ref{TableEoS} for non-rotating
NS. Blue lines represent hadronic EoSs for pure NS. Pink lines are hadronic EoSs
that include hyperons. Green lines are SQM EoSs. The dashed orange line is the
rotational limit imposed by the fastest rotating NS. The dark line in the left
upper corner represents the  causality limit that EoSs must respect. The red band
at the 2 $textrm{M}_\odot$ is the measured value for the J1614-2230 NS. More details in
\cite{2007PhR...442..109L} from where the figure is adapted.
\label{MvsR}}
\end{figure}
The EoS of a neutron star is based on nuclear models that describe nuclear
matter,i.e. a system of interacting nucleons. For the star to be stable, charge
neutrality (total charge resulting from all its constituents) and beta
equilibrium (beta and inverse beta reactions taking place at the same rate) must
hold. Electrons (and at enough high densities muons) form a gas that is
negatively charged. The crust of the star forms a lattice of nuclear clusters
immersed in neutron liquid and as such has solid state properties. The core
behaves like a liquid and presents a mixture of Fermi gases of protons, neutrons
and leptons. In the core, the nuclear energy per baryon is a function of only  
baryon number density $n$ and the isospin asymmetry $\alpha$ and is defined as
follows:
\begin{equation}
E_{nuc}(n,\alpha)=V(n)+S(n)*\alpha^{2}+Q(n)*\alpha^{4}+\mathcal{O}(\alpha^{6})
\end{equation}
where $\alpha \equiv (n_{n}-n_{p})/n$ and $n\equiv n_{p}+n_{n}$ are the neutron
and proton number densities. Here the last term is negligible since it's 
contribution is very small. The most interesting quantity is the Symmetry
Energy (SE) $S(n)$ that has impact in the crust properties of the star.  It's
value  at saturation point (the density of the nucleus of an atom),
$n_{0}=0.16$, is $S(n_{0})=30\pm1$ MeV, as determined by experiments of
isospin diffusion \cite{2005PhRvL..94c2701C} and in agreement with the semi-empirical mass formula
used to describe nuclei. Values at low density have lately been investigated
both in theory and  experiment \cite{2010PhRvL.104t2501N,2007PhRvC..75a4601K}.

The core system is then described by thermodynamic quantities derived from this
energy form (for example $p=-\partial E / \partial V$) at zero Temperature
($T=0$) since its contribution has no effects in the EoS.  To estimate the crust
core transition point one may start by looking for the instability values
against density fluctuations where the system must split into two phases i.e.
where the compressibility of nuclear matter becomes negative,  since the
condition $K_{\mu}=\left(\frac{\partial p}{\partial n}\right)_{\mu} \geq 0 $
must hold. This marks a lower bound in density values where the transition
should occur. The line corresponding $K_\mu=0$ is usually called spinodal line.
 A more elaborated way of addressing this problem is to consider a
first order phase transition of a two component system. The first
consisting only of neutrons while the second having neutrons and protons.
To ensure stability,
mechanical and chemical equilibrium must take place by means of the Gibbs
conditions: 
%\mu^{I}(n)   =  \mu^{II}(n)
%\begin{equation}
% K_{\mu}=\left(\frac{\partial p}{\partial n}\right)_{\mu} \geq 0.
%\end{equation}
\begin{equation}
 p^{I}   =  p^{II} ~~~~~,~~~~~
\mu_n^{I}   =  \mu_n^{II}~~~~~,~~~~~\mu_e^{I}   =  \mu_e^{II}.
\label{coex}
\end{equation}
Protons are present only in second phase, so for them $\mu_p^{I}  > 
\mu_p^{II}$.
In the two component system the average densities and energies are described in terms of
the volume fraction $\chi$ occupied by the ``I'' fluid. They are defined as:
\begin{equation}
 \langle \epsilon \rangle   = \chi\epsilon^{I}+(1-\chi)\epsilon^{II} 
~~~~~~,~~~~~~
 \langle n  \rangle   =  \chi n^{I}+(1-\chi)n^{II}.
\end{equation}
\begin{figure}[t!]
\resizebox{0.8\textwidth}{10cm}
{\includegraphics{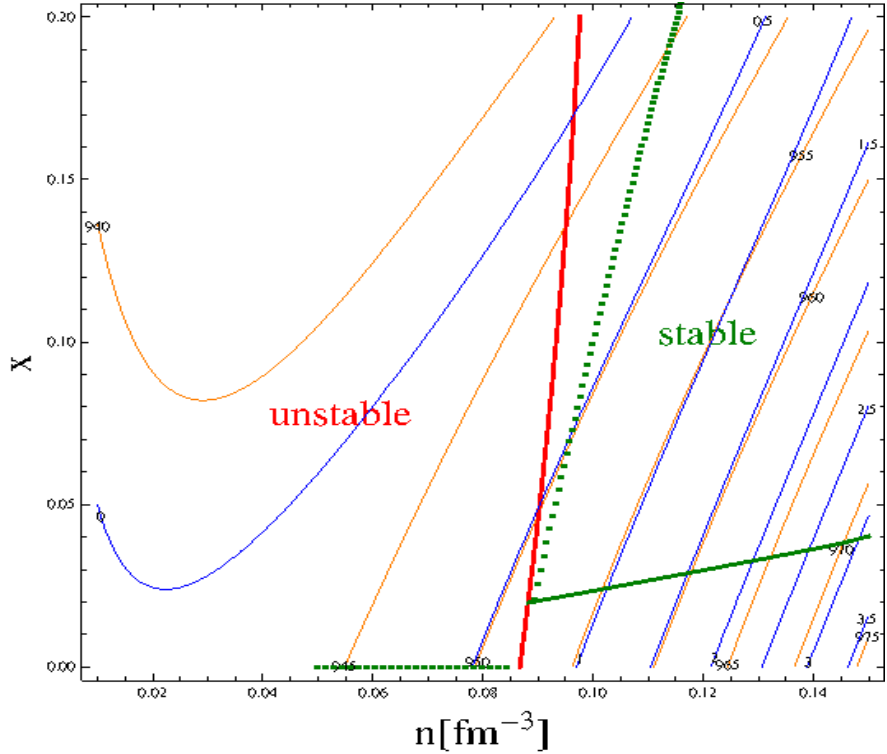}}
\caption{Diagram for  the two component system for the PALu
\cite{1988PhRvL..61.2518P}. Blue contours represent nucleonic pressure while
orange ones neutron chemical potential. The spinodal line is in red. Green dots are
the points where the Gibbs conditions are fulfilled. Points to the left of the
spinodal belong to the pure neutron phase, on the right are points for the
nuclear clusters. The crust-core transition takes place at the density where
the green solid line touches the spinodal. For this model it occurs at
$n_{c}=0.0892\fm3$.}
\label{TwoFluid}
\end{figure}
This energy as a function of density is to be compared with the energy of the
homogeneous system, the transition point being the value for which the  two
component system is preferred, where its energy is the lowest. The behaviour of
matter close to the crust-core transition region is shown in Fig.\ref{TwoFluid}.
Any state of matter is represented by a point on the $x-n$ plane, where $x$ is the
proton fraction $x \equiv n_p/n$ and $n$ is the baryon  number density.
The green solid line  corresponds to the homogeneous neutral matter
filling the NS core. As the density decreases the homogeneous matter becomes
unstable when it crosses spinodal line (the red one). All states  of matter 
on left side of the spinodal are not stable except the pure neutron matter 
($x=0$). That means that for low density the system splits into two phases:
nuclear matter $x\neq 0$ in the form of finite size clusters  immersed in 
pure neutron matter (almost  free neutron + electron gas). Those two phases
are represented by two branches of green points for which the coexistence
conditions (\ref{coex}) hold. 
In this approach the finite size effect like Coulomb and
surface energy are neglected. These different effects lead to formation of 
various structures like rods, plates, and are called pasta phases. See
\cite{kubisAPPB41} for a detailed discussion.

%The low density part of the symmetry energ

\vspace{-.3cm}

\section{Relativistic Measurements}

%Neutron stars are so massive and compact that most of the physical processes around them should be studied in the frame of General Relativity.

Astronomical observations can shed light on NS macroscopic properties and help
to rule out theoretical models. The two basic quantities are mass $M$ and radius
$R$. Fig.\ref{MvsR} shows the expected values for families of stars given an EoS.
Most processes like Eddington flux near the Star's surface, redshift of any
luminous signal and spectra from thermal bursts involve the compactness
parameter $\frac{GM}{Rc^{2}}$ where only the ratio of these quantities is
found \cite{2006Natur.441.1115O,2007Natur.445E...7A}. Nevertheless, combining at least two processes is possible to find
such values, something that has not yet been achieved.  For this task, future 
observations with new extraterrestrial telescopes that have been implemented are
quite promising. In the case of orbiting binary starts only $M$ can be derived
with good accuracy in the case of orbit reduction (by gravitational wave
emission) and  Shapiro delay (radar signal delay near massive objects) lately
measured and resulting in a 2 $\textrm{M}_\odot$ NS \cite{2010Natur.467.1081D}. Macroscopic crust properties
of NS have also been constrained by pulsar glitching: a sudden spin up of the
star. For the case of the Vela pulsar a lower bound  on the crustal moment of
Inertia, $I/I_{crust}\geq 1.4 \% $,  has been established by observations based
on the superfluity model \cite{1999PhRvL..83.3362L}.
\vspace{-.5cm}

\begin{table}[b!] 
\begin{tabular}{lp{4.5in}}
\hline 
Symbol & References and Specifications \\ 
\hline
AkmalPR & Akmal et al.1998 A18+dv+UIX* (npemu) core, BPS+HP94 outer crust, SLy4 inner crust.\\
SLy4 & SLy4 (npemu) core, BPS+HP94 outer crust, SLy4 inner crust, Douchin and Haensel 2001.\\
FPS & BPS below n.drip, then FPS.\\
BPAL12 & Prakash 1997.\\
BalbN1H1 &  BalbN1H1 core, SLy4 crust (S.Balberg,October 1997). \\
GlendNH3 & Nucleons + Hyperons, Lagrangian mean field theory, Glendenning, N.K. 1985 ApJ293, 470 .\\
SQMa & SQM EoS: $\epsilon=p/v^{2}+\epsilon_{b}$, $\epsilon_{b}=5.6\epsilon_{0}$, $v^{2}=1/3$, Glendenning, August 1990.\\
SQMb & SQM EoS: $\epsilon=p/v^{2}+\epsilon_{b}$, $\epsilon_{b}=5.6\epsilon_{0}$, $v^{2}=1$, Glendenning, August 1990.\\
SQMc & SQM EoS: $\epsilon=p/v^{2}+\epsilon_{b}$, $\epsilon_{b}=3.1\epsilon_{0}$, $v^{2}=1/3$, Glendenning, August 1990.\\
\hline
\caption {EoSs used in Fig.\ref{MvsR}}
\end{tabular}
\label{TableEoS}
\end{table}

\section{Summary}

Neutron Stars are dense compact
objects where the most extreme physical conditions exist. The macroscopic
description being in terms of the General Theory of Relativity. For the interior of the star different nuclear
models are being used based on terrestrial laboratory experiments and
extrapolated to the conditions found in NS. Theoretical models should be able to
reproduce observations, and with the forthcoming data such models should start
to converge  since some of them could be tweaked in their parameters to reach the
desired values.  In particular the latest 2 $\textrm{M}_\odot$ measurement imposes a
strong constrain in the EoS.  Finally other relativistic processes like the
cooling of the proto-neutron star could also provide evidence on the interior of
NS, since the cooling rate is modified by the presence of superfluid matter. All
this is reflected in the $M$ vs $R$ relation. \section {Acknowledgments}  We
thank the Gravitation and Mathematical 
Physics Division of the Mexican Physical Society for support and hospitality. This work was
supported by CompStar, a Research Networking Programme of the European Science
Foundation.

%...AkmalPR: Akmal et al.1998 A18+dv+UIX* (npemu)... 
% ...BPS+HP94 outer crust, SLy4 inner crust...

%...SLy4 (npemu)... BPS+HP94 outer crust.....
%# ...SLy4 inner crust... Douchin & Haensel 2001... 

%BPS below n.drip, then FPS

%BPal e.o.s.: crust - SLy ;core: BHF(Paris+TBF)
%  BPAL12 e.o.s.:  . Core: Bombaci August 1999
%Prakash et al. 1997

% BalbN1H1: crust - SLy  core: BalbN1H1
%(S.Balberg,October 1997) 

%Glendenning, N.K. 1985 ApJ293, 470 ......
%nucleons + hyperons .... 
% Lagrangian mean field theory .... 

%EoS: $\epsilon=p/v^{2}+\epsilon_{b}$  Glendenning, August 1990

\newcommand{\aap}{A\&A}
\newcommand{\apj}{ApJ}
\newcommand{\prc}{Phys.~Rev.~C}
\newcommand{\physrep}{Phys.~Rep.~}
\newcommand{\apjl}{ApJ}
\newcommand{\nat}{Nature}

%\vspace{-1cm}

\end{document}